\documentclass{scsPaperFormattingTemplate-LaTex-Revised20150804}
\copyrightnotice{
	SpringSim-HPC 2016 April 3-6, Pasadena, CA, USA
	
	\copyright 2016 Society for Modeling \& Simulation International (SCS) 
}

\usepackage{balance}		
\usepackage{graphics}		
\usepackage{times}			
\usepackage{url}			
\usepackage{dblfloatfix}	

\usepackage{cite}
\usepackage[caption=false,font=footnotesize,labelfont=sf,textfont=sf]{subfig}
\usepackage{algorithm}
\usepackage{algorithmic}
\makeatletter
\def\url@leostyle{%
  \@ifundefined{selectfont}{\def\UrlFont{\sf}}{\def\UrlFont{\small\bf\ttfamily}}}
\makeatother
\def\pprw{8.5in}
\def\pprh{11in}

\usepackage[pdftex]{hyperref}
\hypersetup{
pdftitle={SCS Conference Proceedings Format},
pdfauthor={LaTeX},
pdfkeywords={SCS, proceedings, archival format},
bookmarksnumbered,
pdfstartview={FitH},
colorlinks,
citecolor=black,
filecolor=black,
linkcolor=black,
urlcolor=black,
breaklinks=true,
}

\begin{document}

\title{Server Consolidation for Internet Applications in Virtualized Data Centers}

\numberofauthors{2}
\author{
  \alignauthor Bo Wang\\
    \affaddr{Department of Computer Science and Technology, Xi'an Jiaotong University}\\
    \affaddr{ Xi'an, China 710049 \&}\\
    \affaddr{SKL Computer Architecture, ICT, CAS}\\
    \email{wangbo2012@stu.xjtu.edu.cn}\\
    \alignauthor Ying Song\\
    \affaddr{SKL Computer Architecture, ICT, CAS}\\
    \affaddr{Beijing, China 100190}\\
    \email{songying@ict.ac.cn}\\
    \alignauthor Yuzhong sun\\
    \affaddr{SKL Computer Architecture, ICT, CAS}\\
    \affaddr{Beijing, China 100190}\\
    \email{yuzhongsun@ict.ac.cn}\\
    \alignauthor Jun Liu\\
    \affaddr{SPKLSTN Lab, Department of Computer Science and Technology, Xi'an Jiaotong University}\\
    \affaddr{ Xi'an, China 710049}\\
    \email{liukeen@mail.xjtu.edu.cn}
}

\maketitle

\begin{abstract}
Server consolidation based on virtualization technology simplifies system administration and improves energy efficiency by improving resource utilizations and reducing the physical machine (PM) number in contemporary service-oriented data centers. The elasticity of Internet applications changes the consolidation technologies from addressing virtual machines (VMs) to PMs mapping schemes which must know the VMs statuses, i.e. the number of VMs and the profiling data of each VM, into providing the application-to-VM-to-PM mapping. In this paper, we study on the consolidation of multiple Internet applications, minimizing the number of PMs with required performance. We first model the consolidation providing the application-to-VM-to-PM mapping to minimize the number of PMs as an integer linear programming problem, and then present a heuristic algorithm to solve the problem in polynomial time. Extensive experimental results show that our heuristic algorithm consumes less than 4.3\% more resources than the optimal amounts with few overheads. Existing consolidation technologies using the input of the VM statuses output by our heuristic algorithm consume 1.06\% more PMs.
\end{abstract}


\section{Introduction}

Virtualization technology, such as VM, has been ubiquitously used in cloud computing for resource management. It offers opportunities not only to better isolation and manageability but also to on-demand resource provision for server consolidation. There are many efforts focusing on virtualization, such as resource virtualization, dynamic deployment of virtual machines (VMs), and on-demand resource allocation among the hosted VMs. These works lead to improvements in the performance of virtualization and resource utilizations. Server consolidation based on VM simplifies system administration and improves energy efficiency by improving resource utilizations and reducing the used physical machine (PM) number.


Server consolidation remaps VMs and PMs when the resources needed by the VMs change to minimize the used PM number or energy consumption. While almost all of the existing works \cite{Moldable,TENMIN,IPDPS15,iAware,pMapper,StaticC2,StaticC1,StaticC3,Con1,Con3} addressed the consolidation problem of the applications of which the corresponding VMs are fixed in number at runtime. There are some problems to apply these works to Internet applications, such as e-commerce and web services, whose instances each of which corresponds to exactly one VM can be tuned in number at runtime. The most major one is deciding the number of application instances before consolidation. If the instance number is too large for an application, there would be many underutilized VMs when the load is low, which will increase the used PM number. While if the number is too small, the performance requirement of the application will not be satisfied when the load is high. Thus, the used PM number is affected by not only the resource needed by every VM but also the VM number. So, it should be considered to adjust the number of instances for Internet applications when consolidating them.

Some elasticity managements \cite{HS1,HS2,HS3,HS4,HS5,HS6,AutoC1,AutoC2,CloudScale,VScaler,VS1,VS2,VS3, HV1,HV2,HV3,HV4,HV5,HV6} applied one or the combination of vertical scaling (resizing a VM) and horizontal scaling (adding/deleting VMs) to provide the mapping between elastic applications and VMs leased from a public cloud for minimizing the rent cost in the perspective of cloud users. While they do not consider the placement of VMs on PMs, which is critical for improving energy efficiency. A few of elasticity managements \cite{HV7,HV8,smartSLA} minimized one or more of SLA (service level agreement) penalty cost, rented hardware cost, software cost, and action (e.g. load balancing) cost for service providers. While, all of these works did not take into account energy efficiency which is one major goal of efficient operations in virtualized data centers \cite{DaaC}. Except that, they considered that the VMs with same configuration had identical performance for an application, which is not true for heterogeneous PMs.

In this paper, to our best knowledge, we make the first attempt to consolidate multiple Internet application for improving energy efficiency by minimizing PM number from the perspective of a service provider using its owned cloud. We first model the consolidation providing the application-to-VM-to-PM mapping to minimize the number of PMs as an integer linear programming (ILP) problem, and then present a heuristic algorithm to solve the ILP problem in polynomial time. In brief, the contributions of this paper can be summarized as follows:

\begin{enumerate}
\item We model the consolidation into an ILP problem which provides the application-to-VM-to-PM mapping minimizing PM number guaranteeing required performance.
\item To solve the ILP problem in polynomial time, we propose a heuristic algorithm. Its basic idea is respectively assigning an available PM and the VM instance type both of which provide the best ratios between performance and resource amount to the application with maximum relative difference between required performance and provided performance and then allocating the rest of this PM's resources to applications in the same way.
\item We conduct extensive experiments using various benchmarks to investigate the effectiveness and efficiency of the proposed heuristic algorithm. The experiments results show that our heuristic algorithm consumes only about 4.3\% more resources than the optimal amounts with few overheads and that two existing consolidation technologies using the input of the VM statuses output by our heuristic algorithm consume 1.06\% more PMs.
\end{enumerate}

The rest of the paper is organized as follows. Section \ref{RW} discusses related work. Section \ref{SCF} presents our model and heuristic algorithm. Section \ref{ERA} evaluates our heuristic algorithm and Section \ref{CFW} concludes this paper.

\section{Related Works} \label{RW}
\subsection{VM Consolidation}

The power consumption of a PM when it is powered on but idle is above 50\% of that when it is busy (100\% resource utilization) \cite{Energy}. This motivates server consolidation which increases resource utilizations and energy efficiency by consolidating multiple applications concurrently running on fewer PMs.

Existing server consolidation algorithms \cite{Moldable,StaticC2,StaticC1,StaticC3,Con1,Con3,pMapper,IPDPS15,TENMIN,iAware} provided a target VM-to-PM mapping minimizing the PM number or with other objectives, e.g., VM migration cost and consumed energy, and switched the current mapping to the target mapping by VM migration and resource reallocation (vertical scaling) when some VMs have changes in their required resources. While, to our best knowledge, no existing work has considered consolidating Internet applications whose performance can be tuned by not only vertical scaling but also horizontal scaling for improving energy efficiency. Besides, all of these existing works changed the VM-to-PM mapping by migration leading to non-negligible performance loss and energy overhead \cite{iAware,mig3,PEMforLM11,ProcIEEE14}. In this paper, we consolidate Internet applications taking both vertical and horizontal scaling into account.

\subsection{Elasticity Management}

In a cloud, customers request resources provided by the providers in the form of VMs each of which has a price. Cloud customers pay for the requested resources. Cloud providers should pay for SLA penalty cost, rented hardware cost, software cost and action cost \cite{HV8,smartSLA}. To minimize the cost for cloud customers or providers, a plenty of works \cite{HS1,HS2,HS3,HS4,HS5,HS6, AutoC1,AutoC2,CloudScale, VScaler,VS1,VS2,VS3, HV1,HV2,HV3,HV4,HV5,HV6,HV7,HV8,smartSLA} have studied on scaling the application horizontally and/or vertically.

A few works \cite{HS1,HS2,HS3,HS4,HS5,HS6} studied on the horizontal scaling which tunes the number of VM instances depending to workload variations for an application. Compared to the vertical scaling, horizontal scaling adjusts allocated resources in coarse granularity. Horizontal scaling is supported by most enterprise clouds \cite{EC2}.

Some existing works \cite{AutoC1,AutoC2,CloudScale,VScaler,VS1,VS2,VS3} studied on the vertical scaling which reconfigures VMs. Vertical scaling in comparison to horizontal scaling allows to allocate resources with lower overhead in terms of time and cost \cite{VScaler}. While, vertically scaling up a VM can cause costly migration if its host has no enough resource. Vertical scaling is widely used for dynamic consolidation in data centers \cite{AutoC2,CloudScale,VScaler}.

To more effectively manage the elasticity of applications, some works \cite{HV1,HV2,HV3,HV4,HV5,HV6} combined horizontal scaling and vertical scaling. These works changed the current VM set into the target VM set which provided the required performance with minimal financial expenditure for customers. These works did not take into account the placement of VMs on PMs.

From the service providers' perspective, a few works \cite{HV7,HV8,smartSLA} studied on application-to-VM-to-PM mappings to minimize the cost operating in a per-application level. For example, SmartSLA \cite{smartSLA} horizontally and vertically tuned VMs according to the average SLA penalty cost predicted using machine learning to minimize SLA penalty cost, rented hardware cost, and action cost. Jung et al. \cite{HV7} predicted the behaves of workloads employing an autoregressive moving averages (ARMA) model and then tuned VMs to minimize SLA penalty. 
These works did not take advantage of consolidating multiple applications for improving resource utilizations and energy efficiency.

In addition, all of these above elasticity managements assumed that VMs with same configuration had identical performance for applications, which is not true for heterogeneous PMs.

On the contrary, our work studies on the server consolidation for Internet applications in the perspective of service providers. Our work provides application-to-VM-to-PM mapping to minimize the PM number satisfying the required performance.

\section{Server Consolidation}\label{SCF}



\begin{figure}[!t]
\centering
\includegraphics[width=3in]{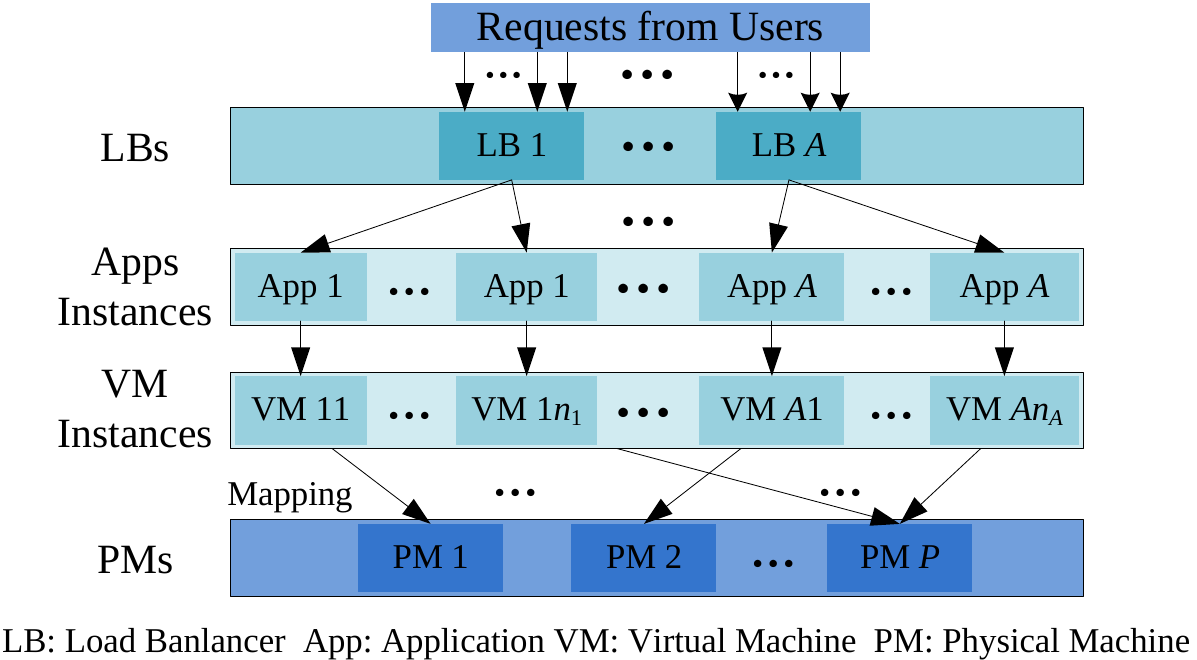}
\caption{The architecture of a virtualized data center providing Internet applications.}
\label{arch}
\end{figure}

In a virtualized data center providing Internet applications, as shown in Fig.~\ref{arch}, a request is distributed to an instance of the corresponding application which has multiple instances each of which is deployed on a VM hosted on a PM, by corresponding load banlancer (LB). The designs of LBs are out of scope of this paper. In this paper, we focus on the server consolidation which minimizes PM number guaranteeing required performance for multiple Internet applications.

In this section, we first present the ILP model of the server consolidation (Section~\ref{Model}) and then describe the heuristic algorithm solving the ILP model in polynomial time (Section~\ref{HA}) in details.

\subsection{Modelling} \label{Model}

The goal of the consolidation is to solve the optimization problem (OP) that is \textit{to minimize the number of used PMs while the provided performance (e.g., throughput) satisfies the requirement for each application}. 

We take $A$ applications, $R$ types of resources, $P$ heterogeneous PMs and $V$ VM configurations ($v_1,...,v_V$) into account. The available amount of resource $j$ ($j=1,...,R$) on PM $k$ is $r_{j,k}$. The required performance of application $i$ ($i=1,...,S$) is $\mu_{i}$. For resource $j$, the configured amount ($l=1,...,V$) is $v_{j,l}$ in VM instance type $l$. On PM $k$, the performance provided by a VM instance with type $l$ for application $i$ is $\mu_{i,k,l}$. We define the variables $x_{i,k,l}$, $i=1,...,S$, $k=1,...,P$, $l=1,...,V$, where $x_{i,k,l}=m$ if there are $m$ VM instances hosted on PM $k$ with type $l$ to provide application $i$, and the binary variables $z_k$, $k=1,...,P$, where $z_{k}=1$ if PM $k$ is used and $z_{k}=0$ if not. Table~\ref{notations} summarizes these notations used in this paper.

\begin{table}[!t]
\renewcommand{\arraystretch}{1.2}
\centering\small
\label{notations}
\begin{tabular}{|c|p{2.3in}|}
\hline
\textbf{Notations} & \textbf{Description} \\\hline
$A$ & The number of Internet applications provided by the hybrid cloud.\\\hline
$R$ & The number of resource types.\\\hline
$P$ & The number of PMs in the private cloud.\\\hline
$V$ & The number of VM instance types.\\\hline
$r_{j,k}$ & The available amount of resource $j$ on PM $k$.\\\hline
$v_{j,l}$ & The amount of resource $j$ configured in VM instance type $l$.\\\hline
$\mu_i$ & The required performance of application $i$.\\\hline
$\mu_{i,k,l}$ & The performance provided by the VM instance with type $l$ on PM $k$ for application $i$.\\\hline
$x_{i,k,l}$ & The variable representing the number of VM instances hosted on PM $k$ with type $l$ for providing application $i$.\\\hline
$z_k$ & The binary variable representing whether PM $k$ is used.\\\hline
$\widehat{\mu_i}$ & The provided performance for application $i$, $\sum_{k=1}^P\sum_{l=1}^V(\mu_{i,k,l}\cdot x_{i,k,l})$.\\\hline
$RPR_i$ & The ratio between the required performance and provided performance for application $i$, $\mu_i/\widehat{\mu_i}$.\\\hline
$R2P_{i,j,k,l}$ & The ratio between the provided performance and the proportion of resource for application $i$ hosted on PM $k$ with VM instance type $l$ for resource $j$, $\mu_{i,k,l}\cdot r_{j,k}/v_{j,l}$.\\\hline
$\overline{N}$ & The average number of VMs hosted on a PM.\\\hline
$RD_i$ & The relative difference between the provided performance and the required performance for application $i$, $\widehat{\mu_i}/\mu_i-1$.\\\hline
\end{tabular}
\caption{Notations.}
\end{table}

We formulate the problem of server consolidation as an ILP problem as follows:

\begin{equation}
\textrm{Minimize}\ \ \sum_{k=1}^Pz_k, \label{objective}
\end{equation}

subject to:
\begin{eqnarray}
\nonumber && \sum_{i=1}^A\sum_{l=1}^V(v_{j,l}\cdot x_{i,k,l})\leq r_{j,k}\cdot z_k,\\
&&\hspace{1.1in} \forall j=1,...,R,\ \forall k=1,...,P,\label{st1}\\
\nonumber && \sum_{k=1}^P\sum_{l=1}^V(\mu_{i,k,l}\cdot x_{i,k,l}) \geq \mu_{i},\\
&&\hspace{2in} \forall i=1,...,A,\label{st2}\\
\nonumber && x_{i,k,l}\geq 0,\textrm{ and is integer},\\
&&\hspace{0.35in} \forall i=1,...,A,\ \forall k=1,...,P,\ \forall l=1,...,V,\label{st3}\\
&& z_k\in \{0, 1\},\ \forall k=1,...,P.\label{st4}
\end{eqnarray}
The decision variables are $x_{i,k,l}$ ($i=1,...,A$, $k=1,...,P$, $l=1,...,V$) and $z_k$ ($k=1,...,P$). The objective (\ref{objective}) of this model is minimizing the PM number. Constraints (\ref{st1}) ensure that the aggregate amount of any resource required by all applications deployed on any PM does not exceed its available amount. Constraints (\ref{st2}) guarantee that the provided performance of any application satisfies the corresponding requirement. Constraints (\ref{st3}) and (\ref{st4}) represent the integrality and binary requirements, respectively, for decision variables. After solving this model, we achieve the application deployments, $x_{i,k,l}$ ($i=1,...,A$, $k=1,...,P$, $l=1,...,V$), and the used PMs, $z_k$ ($k=1,...,P$).

\subsection{The Heuristic Algorithm} \label{HA}

As ILP is NP-hard problem \cite{LIPSUR}, the methods exactly solving the ILP problem, such as enumeration or branch-and-bound, are not feasible to analysing the large scale systems because of their exponential time complexities. Thus, we provide \textit{3MAX} heuristic algorithm to find a near-optimal solution with low overhead.

\begin{algorithm}[!t]\small
\caption{The Heuristic Algorithm}
\renewcommand{\algorithmicrequire}{\textbf{Input:}}
\renewcommand\algorithmicensure {\textbf{Output:} }
\renewcommand{\algorithmiccomment}[1]{{\ /*\textit{#1}*/}}
\begin{tabular}{cp{0.395\textwidth}}
$\mathcal{A}$:& the set of Internet applications, $|\mathcal{A}|$ 2-tuples: (an application, required performance);\\
$\mathcal{P}$:&the set of available PMs, $|\mathcal{P}|$ ($R$+1)-tuples: (an available PM, the available amounts of resource $1,...,R$);\\
$\mathcal{V}$:&the set of available VM instance types, $|\mathcal{V}|$ ($R$+2)-tuples: (an available VM instance type, the configured amounts of resource $1,...,R$, price in public cloud);\\
$\mathcal{PV}$:&the set of the performance of every application running on a VM instance with each type hosted on each PM, $|\mathcal{A}|\cdot|\mathcal{V}|\cdot(|\mathcal{P}|+1)$ 4-tuples: ($a$, $v$, $p$, the performance), where $a\in\mathcal{A}$, $v\in\mathcal{V}$, and $p\in\mathcal{P}$;\\
$\mathcal{M}$:&the set of application deployments, $|\mathcal{M}|$ 4-tuples: (a mapping, $a$, $v$, $p$, the provided performance), where $a\in\mathcal{A}$, $v\in\mathcal{V}$, and $p\in\mathcal{P}$;\\
\end{tabular}
 \\
\begin{algorithmic}[1]\small
\REQUIRE $\mathcal{A}$; $\mathcal{P}$; $\mathcal{V}$; $\mathcal{PV}$
\ENSURE $\mathcal{M}$
\WHILE[$m(i)$ is $i$th element in tuple $m$]{$ (\mathcal{P} \neq \phi) $\\\hspace{0.35in}$\wedge (\exists a)(({a\in\mathcal{A}})\wedge(\sum_{m\in\mathcal{M}\wedge m(2)=a} m(5)< a(2)))$} \label{while}
	\STATE $app \gets a: (a\in \mathcal{A})\wedge $C$1(a) \wedge $C$2(a)$;\\
	/*C$1(a): a(2)>\sum_{m\in\mathcal{M}\wedge m(2)=a}m(5)$;\\
	C$2(a):\frac{a(2)}{\sum\limits_{m\in\mathcal{M}\wedge m(2)=a}m(5)} = $\\\hspace{1in}$\max\limits_{a'\in \mathcal{A}\wedge \textrm{C}1(a')}\frac{a'(2)}{\sum\limits_{m\in\mathcal{M}\wedge m(2)=a'}m(5)}$;*/
	\STATE $pm\gets p: (p\in \mathcal{P})\wedge $C$3(p) \wedge $C$4(p)$;\\
	/*C$3(p): R2P(app, p) = \max\limits_{p'\in\mathcal{P}}R2P(app, p')$;\\
	$R2P(a, p)=\max\limits_{\substack{(pv\in\mathcal{PV}) \wedge (pv(1)=a)\wedge (pv(3)=p)\\\wedge(v\in\mathcal{V})\wedge (pv(2)=v)\wedge(2\leq j \leq R+1)}} \frac{pv(4)\cdot p(j)}{v(j)}$;\\
	C$4(p): (\forall j \in \{2,...,R+1\}) (p(j) = \max\limits_{(p'\in \mathcal{P})\wedge \textrm{C}3(p')}{p'(j)})$;*/
	\STATE $\mathcal{P} \gets \mathcal{P}\setminus pm$;
	\WHILE{true}
		\STATE $app \gets a: (a\in \mathcal{A})\wedge $C$1(a) \wedge $C$2(a)$;
		\STATE $vm \gets v: $C$5(v) \wedge $C$6(app, v, pm)\wedge $C$7(app, v, pm)$;\\
		/*C$5(v): (v\in \mathcal{V})\wedge ((\forall j \in \{2,...,R+1\})(v(j)\leq pm(j)))$;\\
		C$6(a, v, p): pv(4)|_{(pv(1)=a)\wedge (pv(2)=v)\wedge (pv(3)=p)}=$\\\hspace{0.6in}$ \max\limits_{\textrm{C}5(v')}pv(4)|_{(pv(1)=a)\wedge (pv(2)=v')\wedge (pv(3)=p)}$;\\
		C$7(a, v, p): $\\\hspace{0.1in}$(\forall j \in \{2,...,R+1\})(v(j) = \min\limits_{\textrm{C}5(v')\wedge \textrm{C}6(a, v', p)}v'(j))$;*/
		\IF {$vm = null$}
			\STATE \textbf{goto} line \ref{while};
		\ENDIF
		\STATE $pv \gets pv': (pv'\in \mathcal{PV})$\\\hspace{0.3in}$\wedge(pv'(1)=app)\wedge (pv'(2)=vm)\wedge (pv'(3)=pm)$;
		\STATE $\mathcal{M} \gets \mathcal{M} \cup \{($a new mapping$, pv)\}$;
	\ENDWHILE
\ENDWHILE
\end{algorithmic}
\end{algorithm}

The basic idea of \textit{3MAX} is selecting the application with maximum ratio between the required performance and provided performance ($RPR_i = \mu_{i}/\sum_{k=1}^P\sum_{l=1}^V(\mu_{i,k,l}\cdot x_{i,k,l})$) and respectively assigning PM and the VM instance type both of which provide best (maximal) performance to the application. This is why we call it \textit{3MAX}. The details, outlined in Algorithm 1, are presented as follows.

\textit{Step} 1. \textit{3MAX} selects a PM with available resources. For each type of resource, there is a ratio ($R2P_{i,j,k,l}=\mu_{i,k,l}\cdot r_{j,k}/v_{j,l}$) between the provided performance and the proportion of resource on a PM for each VM instance type when providing the application with maximum $RPR_i$. \textit{3MAX} selects the PM with the maximum of these ratios (lines 2-4). If there are multiple PMs giving the maximum ratio, \textit{3MAX} selects the PM with most amount of available resources from these PMs (C4 in line 3).

\textit{Step} 2. \textit{3MAX} allocates a VM to the application with maximum $RPR_i$ (line 6) on the PM selected in \textit{Step} 1. \textit{3MAX} allocates a VM with the type giving the best performance for this application (line 7). If there are multiple VM instance types giving the best performance, \textit{3MAX} selects the VM instance type configured minimal amounts of resources (C7 in line 7).

\textit{Step} 3. \textit{3MAX} repeats \textit{Step} 2 until there is no available resource on the selected PM for any application (lines 8-10).

\textit{Step} 4. \textit{3MAX} repeats \textit{Step} 1-3 until the provided performances of all applications are satisfying their respective requirements or there is no available PM (line 1).

The time complexities of the selections (\textit{Step} 1-2) of application (lines 2 and 6), PM (line 3), and VM instance type (line 7) are $O(A)$, $O(P)$ and $O(V)$, respectively. We assume that a PM hosts $\overline{N}$ VMs on average, then allocating the available resources of a PM (\textit{Step} 1-3, lines 2-13) is $O(APV\overline{N})$ in time complexity. Thus the time complexity of \textit{3MAX} is $O(AP^2V\overline{N})$ at worst.

\section{Performance Evaluation}\label{ERA}

In this section, we introduce our testbed and experiment design. And then we discuss the experimental results.

\subsection{Testbed and Experiments Design}\label{TED}

\begin{table}\small
	\centering
    \renewcommand{\arraystretch}{1.1}
    \begin{tabular}{|c|c|c|}
        \hline
        \textbf{CPU (\#cores)}&\textbf{MEMORY}&\textbf{\#PM}\\ 
        \hline
        Intel(R) Xeon(R) CPU E5410	&8GB	&10\\
        	  @ 2.33GHz (8)		&	&	\\\hline
        Quad-Core AMD Opteron(tm)	&8GB	&10\\
			Processor 2378 (8)	&	&	\\ \hline
        Dual-Core AMD Opteron(tm)	&4GB	&10\\
        	Processor 2216 (4)	&	&	\\\hline
    \end{tabular}
    \caption{The configuration of PMs for hosting VMs.}\label{tb1}
\end{table}

In our testbed, the configurations of PMs used as servers for applications are shown in Table~\ref{tb1}. Each PM is configured with two 1000Mbps Network Interface Cards (NICs).

We select five applications from four benchmarks for our experiments, as follows.

\begin{itemize}
	\item \textbf{TPC-W} \cite{tpcw} is a transactional web e-Commerce benchmark. The specification defines three different mixes of web interactions, each varying the ratio of browse to buy activities. We use TPC-W with the options, the TPC-W Shopping Mix and 1.0 think time, to generate workloads. The performance is measured by Web Interactions Per Second (WIPS).
    \item \textbf{\textit{Yahoo! Cloud Serving Benchmark}} (\textbf{YCSB})\cite{ycsb} is a performance measurement framework for cloud serving systems. Six core workloads (Workload A-F) are provided. A tool YCSB Client is developed to execute the YCSB benchmarks. We chose Workload A, which has 50 percent reads and 50 percent updates, as the workload generator. One million records are loaded into each database server. Performance is evaluated by throughput (operations per second).
    \item \textbf{Apache Benchmark} (\textbf{ab}) \cite{ab}  is a tool for benchmarking HTTP server. We design two applications by the benchmark, \textit{\textbf{abk}} and \textit{\textbf{abm}}. They transmit a fixed size file: 1KB, 1MB, which are representative log sizes in current data center \cite{logsize}, to their requests, respectively.
Additional, to reduce the disk readings for increasing the performance, files are cached in the buffers in advance. Performance is evaluated by throughput (finished requests per second).
    \item \textbf{SysBench} \cite{sysbench} is a modular, cross-platform and multi-threaded benchmark tool for evaluating operate system (OS) parameters. We use the CPU performance benchmark which is one of the most simple benchmarks in SysBench. In this mode each request consists in calculation of prime numbers up to a specified value (20000 in the paper). Events (i.e. finished requests) per second (EPS) is the performance metric.
\end{itemize}

The configurations for VM instances have 4, 4, and 5 options for CPU (1-4 virtual CPUs (VCPUs)), memory ((1-4)$\times$0.5GB), and NIC ((1-5)$\times$200Mbps), respectively. Thus there are 80 different VM instance types. We do not consider the disk resource because the disk resource is never the bottleneck in all of our experiments. We assume that SysBench is an Internet application for which its processing results are the data returned to users for the requests.

We use the trace collected from the 1998 World Cup Web site \cite{WC98} at the five days from May 28, 1998 to June 1, 1998 to generate the workloads of the five applications, respectively, in the following experiments. For an application, we scale the average request number per second within 15 minutes of the trace data by a factor so that the maximal scaled value is equal to the maximum throughput, $mt_i$, which is two fifths of the value of aggregating throughputs provided by all PMs when running the application on a VM with type provisioning best performance, $mt_i=\frac{2}{5}\sum_{k=1}^P\max_{1\leq l\leq V}\mu_{i,k,l}$, and set the scaled values as the required throughputs. Our consolidation algorithm, \textit{3MAX}, runs every 15 minutes.

In the following experiments, we pin each VCPU of hosted VMs on a CPU core because performance loss of virtualization can be reduced by core pinning. On a server, the aggregated number of VCPUs of all hosted VM is not larger than the number of CPU cores to avoid the additional overhead of overcommitment.

Next, we first evaluate the performance of our heuristic algorithm on the accuracy (Section~\ref{Accuracy}), minimizing PM number (Section~\ref{PMnumber}) and scalability (Section~\ref{Scalability}), and then experimentally study on the sensitivity of our algorithm on the accuracy of workload evaluations (Section~\ref{sens}).

\subsection{Accuracy} \label{Accuracy}

We measured the performance of \textit{3MAX} using the relative differences between the total throughputs ($\widehat{\mu}_i$) achieved by \textit{3MAX} and the required throughputs ($\mu_i$), $RD_i = (\widehat{\mu}_i-\mu_i)/\mu_i, i=1,...,A$, and the overall relative difference, $ORD = \frac{1}{A} \sum_{i=1}^A(\sum_t|RD_i|/t)$,  where $t$ represents the experiment time intervals. The closer to 0 $RD_i$ and $ORD$ are, the less resources are wasted and thus the better our algorithm performance is, when $RD_i\geq 0,\ i=1,...,A$. $RD_i<0$ indicates that the required performance is not satisfied for application $i$.

\begin{figure}[!t]
    \centering
    \includegraphics[width=0.8\columnwidth]{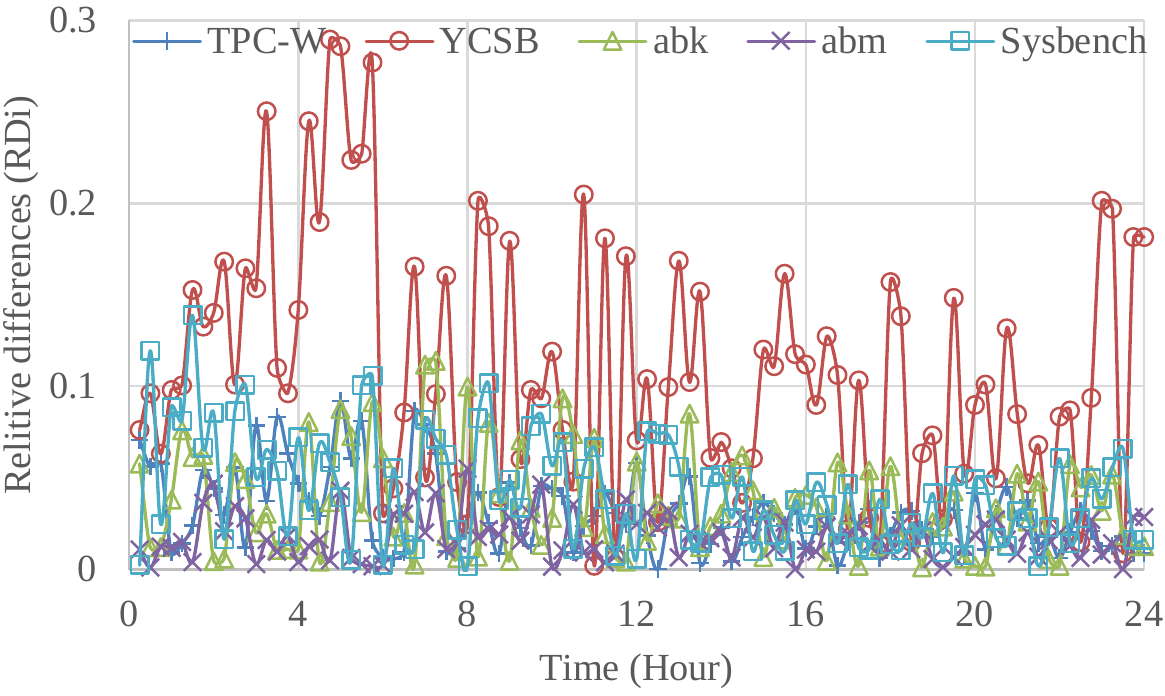}
    \caption{{The performance of our consolidaton algorithm.}}\label{rdct}
   \vspace{-4mm}
\end{figure}

Figure~\ref{rdct} shows the relative differences of applications, provided by our consolidation algorithm. As shown in this figure, $RD_i, i=1,...,A$ are all close to 0, which means that our consolidation algorithm always has a high accuracy all the time. $RD_i, i=1,...,A$ are always slightly larger than 0, which means that the performance achieved by the consolidation is always satisfying the required performance, guaranteed by the termination condition that all required performances of applications are satisfied (line 1 in Algorithm 1). The $ORD$ is about 0.043, i.e., using our consolidation algorithm consumes only about 4.3\% more resources than the optimal amounts.

\subsection{Performance in Minimizing PM Number}\label{PMnumber}

In this section, we experimentally study on the performance of \textit{3MAX} in minimizing PM number by two existing consolidation algorithms. We respectively run these consolidation algorithms with the VM statuses output by \textit{3MAX}. If the PM number is not reduced by these algorithms, then PM number is minimized by \textit{3MAX} in practice is proved. These two consolidation algorithms respectively are First Fit Decreasing (FFD) \cite{pMapper} and Least Loaded (LL) \cite{sca}. The commonly used FFD packing algorithm places the largest VM on the first physical server on which it will fit. If there is no such server, the VM is placed in a new empty server. Least loaded (LL) algorithm \cite{sca} assigns the current VM to the used PM with least-load or a new server when there is no room for the VM in the used PMs. When running these algorithms in our experiment, we sort the PMs by the amount of resources in descending order in advance.

\begin{figure}[!t]
    \centering
    \includegraphics[width=0.8\columnwidth]{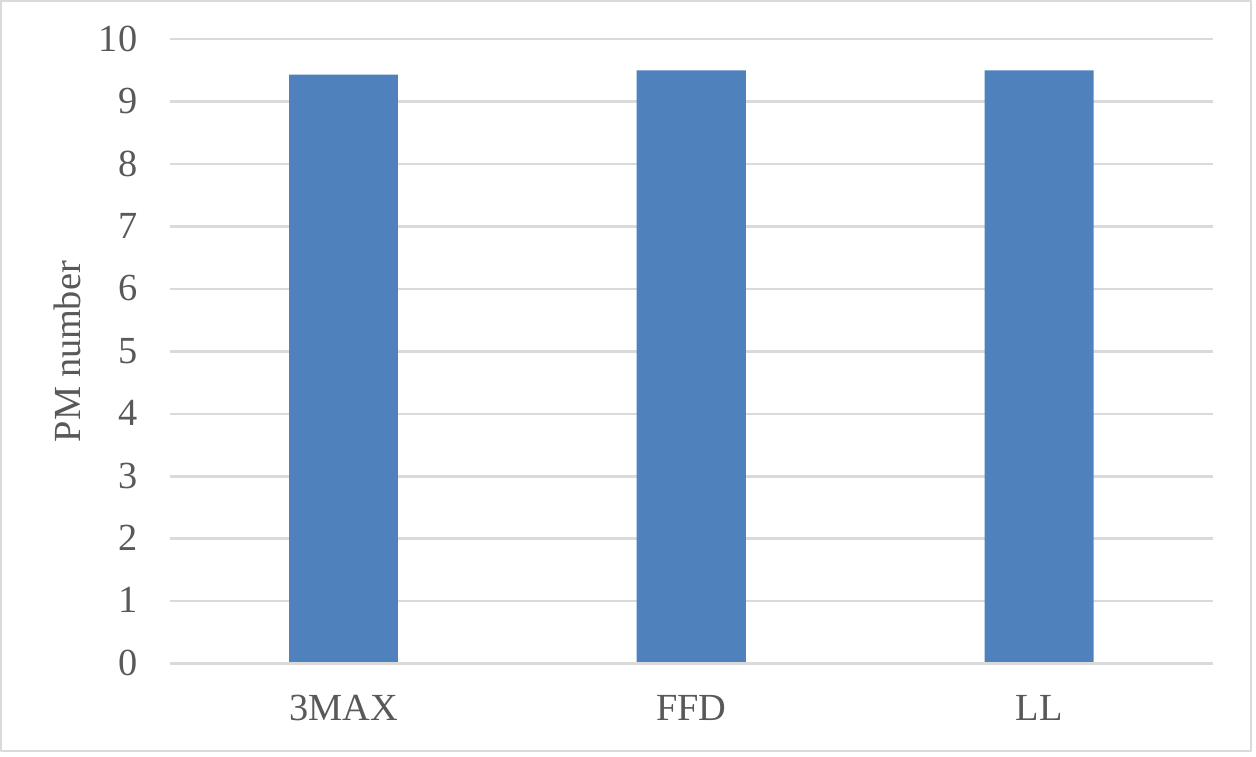}
    \caption{{The average PM numbers consumed by various consolidation algorithms.}}\label{comparison}
   \vspace{-4mm}
\end{figure}

The results are presented in Fig.~\ref{comparison}. From Fig.~\ref{comparison}, we can see that FFD and LL have the same performance, consuming 9.5 PMs on average. \textit{3MAX} consumes about 9.4 PMs whose number is less than that consumed by FFD and LL, on average. That is to say, these two consolidation technologies need about 1.06\% more PMs. Thus, after deploying VMs whose placement is output by \textit{3MAX}, the PM number could not be reduced by these consolidations.

\subsection{Scalability} \label{Scalability}

\begin{figure}[!t]
    \centering
    \subfloat[Scalability in PM number]{
    	\includegraphics[width=0.45\columnwidth]{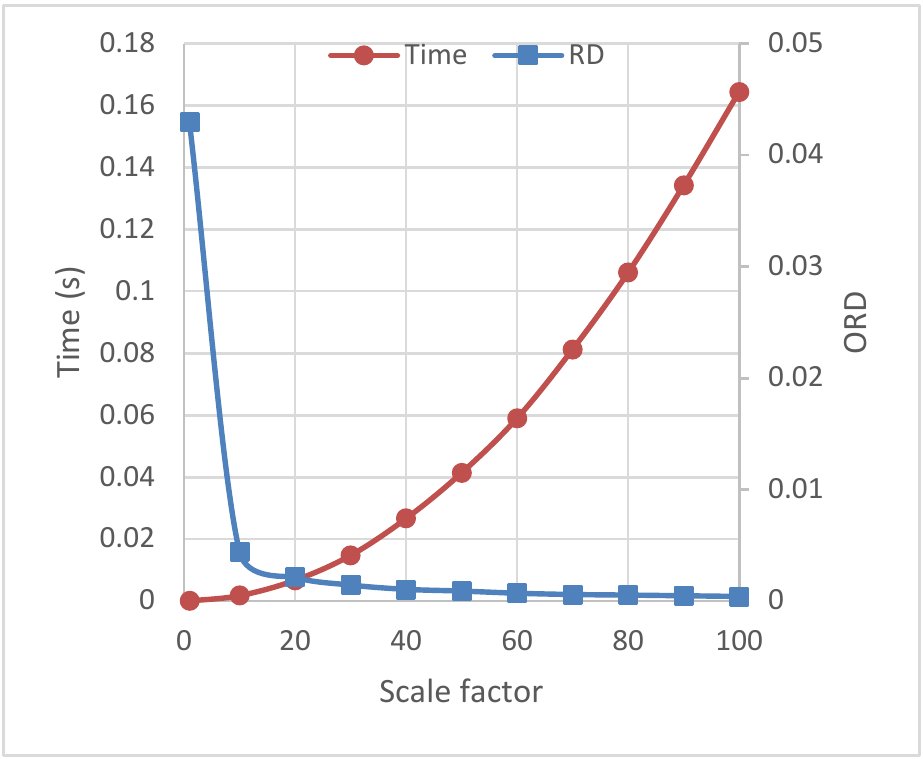}
    }
    \subfloat[Scalability in application number]{
    	\includegraphics[width=0.45\columnwidth]{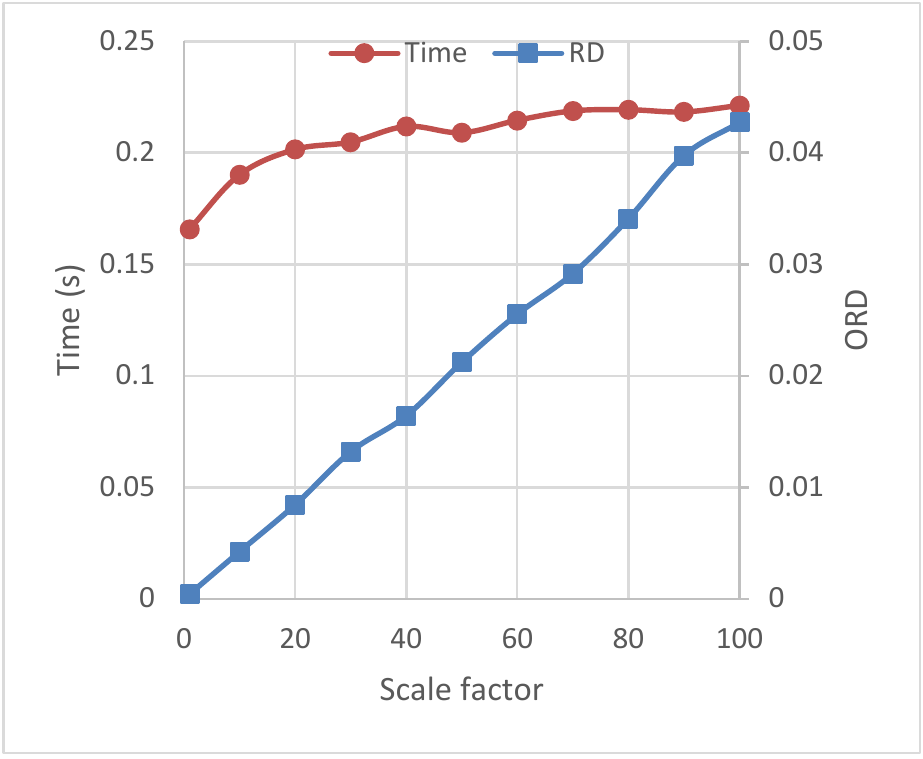}
    }
    \caption{{Scalability of \textit{3MAX} in the numbers of PMs and applications.}}\label{scale}
    \vspace{-4mm}
\end{figure}

In this section, we evaluate the scalability of \textit{3MAX} in consumed CPU time and ORD. We first scale the PM number by a factor ($f_{PM}$) ranging from 1 to 100 to examine the scalability as the PM number increases, and then scale the application number in the same way ($f_{APP}$) with 3000 PMs to study on the scalability as the number of applications increases. For example, \textit{3MAX} consolidating 5 applications on 30 PMs when $f_{PM}=1$, same as the original system decreased in Section~\ref{TED}, and consolidating 5 applications on 3000 PMs when $f_{PM}=100$. We scale the required throughputs of applications in original system by the factor, the ratio of the numbers of PMs and applications, and set them as corresponding required throughputs in the scaled system. We use the average application performance measured in the original system as that in the scaled systems. The results of running on a Quad-Core AMD Opteron(tm) Processor 2378 core are respectively shown in Fig.~\ref{scale}.

Figures \ref{scale}a and \ref{scale}b show a pattern of ORD change, ORD decreases with increasing the workloads of applications. The reasons are as follows. We consider that the resources are allocated to the VMs on a PM in a discrete way, such as CPU is allocated in a granularity of cores, as done in most of clouds, such as EC2 \cite{EC2} and OpenStack \cite{openstack}. The absolute difference between provided performance and required performance is less than the performance provided by a VM with minimal resources using \textit{3MAX}. Therefore, the relative difference is decrease with increasing the workload for an application.

As shown in Fig.~\ref{scale}a, the time consumed by \textit{3MAX} increases quadratically with the PM number, which is consistent with the analysis in Section \ref{HA}. As the application number increases, the consumed time, as shown in Fig.~\ref{scale}b, increases slightly because the consumed time depends largely on the PM number due to that the number of applications are much less than that of PMs. Our algorithm can make a decision only about 0.22 seconds which is much less than the decision-making periods (tens or hundreds of seconds) in most of clouds, even in the case of consolidating 500 applications on 3000 PMs.

\subsection{Sensitivity} \label{sens}

In this section, we experimentally study on the impact of the accuracy of workload evaluation on the accuracy and fluctuation of our consolidation algorithm. We use two workload evaluation methods, $\chi^2$ and $F$. These methods first test whether the workload of an application is changed by $\chi^2$-test \cite{Statistics} and $F$-test \cite{Statistics}, respectively, using the data of current time window containing the time intervals of last requests, and evaluate the workload as the average value within current time window if the workload is tested to be changed. We set the size of time windows and the level of significance of tests as 1000 and 0.01, respectively.

As $F$-test has the assumption that the time intervals of requests follow exponential distribution \cite{Statistics}, we generate workloads for evaluating workload evaluation methods and their impact on our consolidation algorithm as follows. We first generate 100
random numbers and realignment them as 20 combinations each of which contains 5 numbers respectively set as the workloads of the five applications. Then, we sample the exponential distributions with means of the generated 100 random numbers and take the samples as the input of evaluation methods. We set that each combination lasts 100 seconds.

\begin{figure}[!t]
    \centering
    \includegraphics[width=0.8\columnwidth]{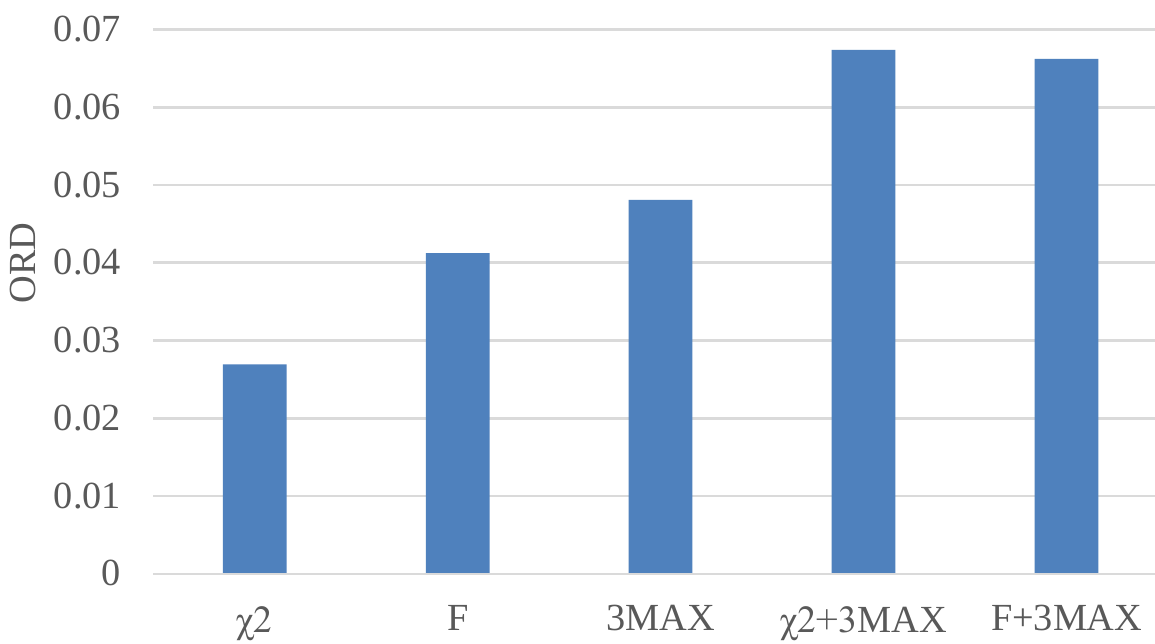}
    \caption{{The impact of the workload evaluation methods on \textit{3MAX}.}}\label{rds}
   \vspace{-2mm}
\end{figure}

The accuracy of workload evaluation is evaluated by the overall relative difference between the estimated workload $\widehat{\lambda}$ and the real workload $\lambda$, $ORD = \sum_t{(|\widehat{\lambda}-\lambda|/\lambda)}/t$. The closer to 0 $ORD$ is, the more accuracy an evaluation method is. Figure~\ref{rds} shows the overall relative differences of these two evaluation methods and our consolidation algorithm respective using the workloads evaluated by these two methods.

As shown in Fig.~\ref{rds}, ORDs of $\chi^2$ and $F$ are respectively 0.027 and 0.041, which degrade 0.043 of the ORD of \textit{3MAX} into 0.0674 and 0.0662, respectively, i.e., these workload evaluation methods have only a little influence on the accuracy of our consolidation algorithm.

\begin{figure}[!t]
    \centering
    \includegraphics[width=0.8\columnwidth]{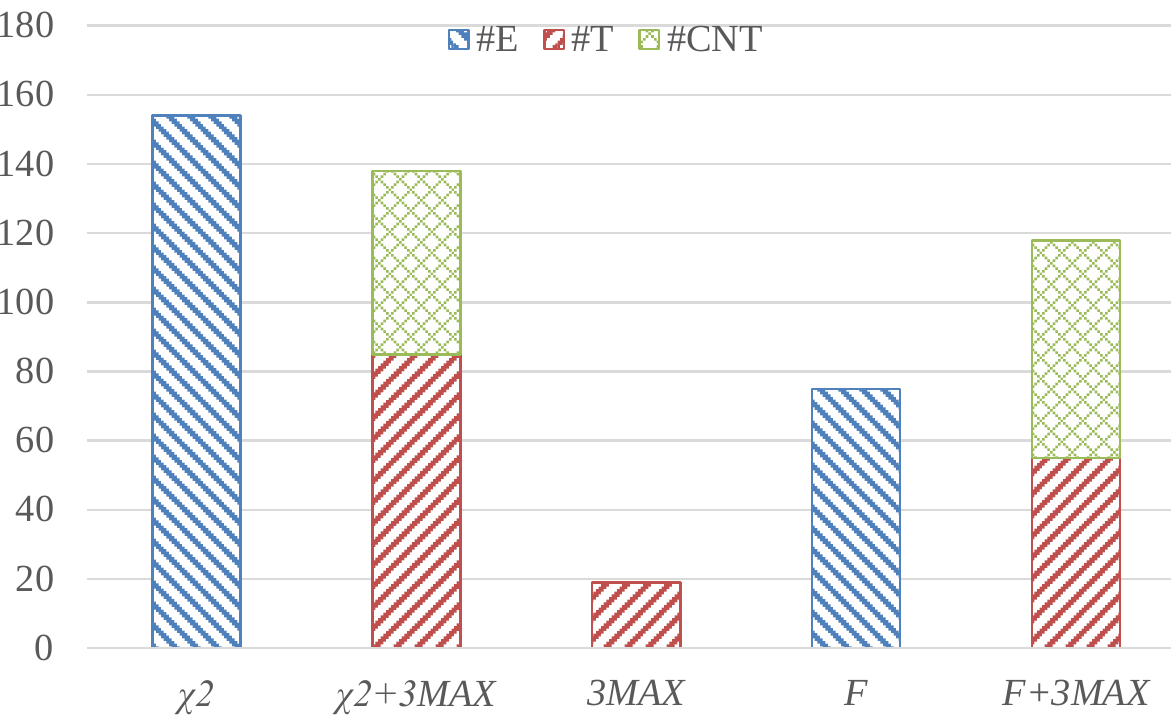}
    \caption{{The numbers of evaluated workload changes (\#E), application deployment changes (\#T), and consolidations without changing deployment (\#CNT).}}\label{ECT}
   \vspace{-4mm}
\end{figure}

The workloads of the five applications change 19 times, while these two evaluation methods respectively change the evaluated workloads 154 and 75 times, all of which are much more than the actual value, as shown in Fig.~\ref{ECT}. This is because the workloads would be re-evaluated even when only one workload are judged to change by the test, which dramatically increases the fluctuations of the evaluation methods. Our consolidation algorithm reduces the fluctuations in two ways. The first is that two combinations of the workloads which have a small difference may correspond to a same application deployment, which reduces the number of application deployment changes. The second is that the current deployment may still satisfy the requirement when the evaluated workloads have small changes, which reduces the number of consolidations. Thus the numbers of application deployment changes, 85 and 55, shown in Fig.~\ref{ECT}, are only about half of the numbers of evaluated workload changes, respectively.

\textit{3MAX} alleviates fluctuations of evaluation methods while does not eliminates them. The deployment change numbers of $\chi^2$+\textit{3MAX} and $F$+\textit{3MAX} are more than that of \textit{3MAX}, as shown in Fig.~\ref{ECT}. Thus, more accuracy of workload evaluation method helps \textit{3MAX} to be more practical.

\section{Conclusion}\label{CFW}

In this paper, to our best knowledge, we make the first attempt to study on consolidating multiple Internet applications in virtualized data centers. We model the consolidation into a ILP problem providing a three-tiered mapping, application-to-VM-to-PM, to minimize the PM number satisfied the required performances of applications. To solve the ILP problem in polynomial time, we propose a heuristic algorithm. Extensive experiments have been conducted to study on the effectiveness and efficiency of our heuristic algorithm. The experiments results show high accuracy of our heuristic algorithm having little sensitivity to the accuracies of workload evaluation methods with good scalability.

\section*{ACKNOWLEDGMENT}
The authors are grateful to the anonymous reviewers' comments. This work was supported in part by the project of NSFC under grants 61202060, 912183001, 61173112 and 61221062, the National High-Tech Research and Development Program (863) of China under grant 2013AA01A212.

\small
\bibliographystyle{unsrt}
\bibliography{VM-SpringSim-new}
\end{document}